\documentclass[12pt,letterpaper]{article}

\usepackage{graphicx,array}
\usepackage{url}
\usepackage{color}
\usepackage{latexsym}
\usepackage{amsthm}
\usepackage{amsmath}
\usepackage{amssymb}
\usepackage{amsfonts}
\usepackage[numbers,sort&compress]{natbib}
\usepackage{bm}
\usepackage{slashed}
\usepackage{mathrsfs}
\usepackage{enumerate}
\usepackage{tikz}
\usepackage{siunitx}
\usepackage{mdframed}
\usepackage{setspace}  
\usepackage{esvect}
\usepackage{esint}
\usepackage{subcaption}
\usepackage[normalem]{ulem}
\usepackage{setspace}
\usepackage{multirow}
\usepackage{diagbox}
\usepackage{soul}
\usepackage{multirow}

\usepackage[version=4]{mhchem}

\usepackage[utf8x]{inputenc}%
\usepackage{tcolorbox}%
\usepackage[normalem]{ulem} 

\usepackage{hyperref} 
\hypersetup{
    colorlinks=true,       
    linkcolor=red,          
    citecolor=blue,        
    filecolor=magenta,      
    urlcolor=blue           
}
\usepackage[all]{hypcap} 

\usepackage{natbib}
\newcommand{\nc}{\newcommand}

\nc{\beq}{\begin{equation}}  
\nc{\eeq}{\end{equation}}  
\nc{\beqa}{\begin{eqnarray}}  
\nc{\eeqa}{\end{eqnarray}}  
\nc{\bit}{\begin{itemize}}  
\nc{\eit}{\end{itemize}}

\usepackage{floatrow}
\newfloatcommand{capbtabbox}{table}[][\FBwidth]

\usepackage{blindtext}

\def\figureautorefname~#1\null{Fig.\,#1\null}
\def\tableautorefname~#1\null{Tab.\,#1\null}

\def\equationautorefname~#1\null{Eq.\,(#1)\null}

\title{
{\bf Near-Extremal Charged Black Holes: \\{\Large Greybody Factors and Evolution}}
	\author{\large Yang Bai and Mrunal Korwar}
	\date{\small \it 
	Department of Physics, University of Wisconsin-Madison, Madison, WI 53706, USA\\
	    }
}

\begin{document}

\maketitle

\setlength{\parskip}{0.2ex}

\begin{abstract}	
As a charged black hole reaches its extremal state via Hawking radiation, quantum effects become important for its thermodynamic properties when its temperature is below a mass gap scale. Using AdS$_2$/CFT$_1$ correspondence and solutions for the corresponding Schwarzian action, we calculate the black hole greybody factors including the quantum effects. In the low temperature limit, the greybody factors scale as $T^{2s + 3/2}$ with $s$ the radiated field spin. Hence, the Hawking radiation of a near-extremal charged black hole (NEBH) is dominated by emitting scalar particles including the Higgs boson. Time evolution of an NEBH is also calculated and shows a stochastic feature. For an NEBH lighter than around $10^8$ times the Planck mass, its temperature at the current universe is below the mass gap scale and is universally tens of GeV, which is important if one searches for primordial (hidden) charged black holes. 
\end{abstract}

\thispagestyle{empty}  
\newpage  
  
\setcounter{page}{1}  


\section{Introduction}\label{sec:Introduction}

The charged or Reissner--Nordstr\"{o}m black hole is one of the simplest solutions in general relativity for a black hole with mass ($M$) and charge ($Q$).
As pointed out a while ago, the semiclassical description for statistical or thermal properties of a charged black hole breaks when it reaches the deep near-extremal limit~\cite{Preskill:1991tb,Maldacena:1998uz}. The black hole energy, defined as $E \equiv M - M_{\rm e}$ with $M_{\rm e}$ as the extremal mass proportional to $Q$, becomes comparable to its temperature such that it is ambiguous to use the temperature before or after the Hawking radiation. This happens when the temperature is below around a mass gap scale $\Lambda_{\rm gap} \equiv M_{\rm pl}^4/M_{\rm e}^3$ with $M_{\rm pl} = 1.22\times 10^{19}$~GeV as the Planck mass. When $T \lesssim \Lambda_{\rm gap}$, additional quantum fluctuations of the low-energy degrees of freedom become important and should be taken into account to describe the black hole thermodynamics. Owing to the universal behaviors of the near-horizon region of a near-extremal charged black hole (NEBH), the quantum effects have been incorporated recently into the black hole statistical properties~\cite{Iliesiu:2020qvm}. The near-horizon geometry of an NEBH is given by ${\rm AdS}_{2}\times S^{2}$, which after dimensional reduction, leads to 2-dimensional dilaton gravity known as Jackiw-Teitlboim (JT) gravity~\cite{Jackiw:1984je,Teitelboim:1983ux}. The bulk JT action can be described entirely in terms of a Schwarzian theory on its boundary~\cite{Almheiri:2014cka, Maldacena:2016upp,Nayak:2018qej} (also see \cite{Mertens:2022irh} for a recent review on the topic). The relevant degree of freedom is the ``Goldstone boson" related to the coset space of time reparametrization symmetry over $SL(2, \mathbb{R})$~\cite{Maldacena:2016upp} or $\mbox{diff}(S^1)/SL(2,  \mathbb{R})$~\cite{Stanford:2017thb}. The path integral involving the Schwarzian action can be solved exactly, leading to quantum corrections to the entropy and the energy of an NEBH~\cite{Stanford:2017thb,Yang:2018gdb, Iliesiu:2019xuh}. The quantum-corrected relation between energy and temperature is given by $E=2\pi^2\,T^2/\Lambda_{\rm gap} \,+\, 3\,T/2$ with the last term coming from quantum effects and more important than the first term when $T \lesssim \Lambda_{\rm gap}$. The density of states of an NEBH is continuous below the mass gap scale, which implies that there is no real mass gap between the black hole extremal state and its near-extremal excitations~\cite{Sachdev:2019bjn,Iliesiu:2020qvm}.

Given the recent progress on understanding the statistical properties of NEBH, we calculate the greybody factors of a charged black hole including the quantum effects in this paper. Both the modified energy-temperature relation and the greybody factors will affect the black hole evolution via Hawking radiation. 
It is well known that in the semiclassical region the greybody factors can be calculated by considering scattering of the matter fields in the black hole background and evaluating the transmission probabilities~\cite{Teukolsky:1973ha, Page:1976df,Maldacena:1997ih, Cvetic:1997ap, Arbey:2019mbc}. For an NEBH, the results can also be obtained by calculating the thermal two-point function of boundary operators in the holographic dual conformal field theory (${\rm CFT}_{d}$) of the near-horizon (${\rm AdS}_{d+1}$) gravity background~\cite{Maldacena:1997re,Gubser:1997cm, Aharony:1999ti}. For the 4-dimensional NEBH and after dimensional reduction to the AdS$_2$ gravity theory, the dual theory is described by the Schwarzian action. The thermal Green's functions of boundary operators in the Schwarzian theory have been evaluated recently in Refs.~\cite{Mertens:2017mtv, Kitaev:2018wpr}. Utilizing their results, we derive a formula for greybody factors of radiating  a field with an arbitrary scaling dimension $\Delta_{\mathcal{O}}$ for its dual operator in the Schwarzian theory. Our greybody factor formula in Eq.~\eqref{eq:gammafac-full} includes the quantum effects and is expressed as a function of the black hole temperature $T$, the mass gap scale $\Lambda_{\rm gap}$ and the radiated quantum frequency $\omega$. In the semiclassical region with $T\gg\Lambda_{\rm gap}$, our formulas match the existing formulas in literature~\cite{Page:1976df,Cvetic:1997ap}. For a lower temperature with $T < \Lambda_{\rm gap}$, the quantum effects become important and the greybody factors show a significant dependence on spins of the radiating fields. The greybody factor is larger for a radiating field with a smaller corresponding operator scaling dimension. To calculate the operator scaling dimensions, we perform a dimensional reduction for various fields in the appendix including a massive spin-1 boson, which has not been done in the literature, but is important for massive weak gauge boson radiation. The greybody factor for radiating a spin-0 boson (including the ``scalar" degree of freedom of a massive gauge boson) is found to be much larger than radiating other particles. 

After knowing the greybody factors, we can then calculate the time evolution of an NEBH at least including the Standard Model (SM) particles as the Hawking radiation fields. Interestingly, we have found that the Higgs boson or the electroweak boson sector plays an important role. Because of the relation $E \approx 3\,T/2$ in the quantum-dominant region, the NEBH time evolution follows a stochastic way. After the black hole radiates a quantum $\omega$, its temperature becomes $2(E-\omega)/3$, which is order-one different from previous step. Because of power-law dependence in $T$ for the greybody factors, the black hole takes a longer time to emit the next quantum with a smaller energy. The black hole emission is more Poisson-like with discretized changes in energy and temperature. On the other hand, the statistic features of a large ensemble of black holes can be simulated, which we will show.

The rest of this paper is organized as follows. In Section~\ref{sec:property}, we describe some basic properties of NEBH, specifically the black hole energy-temperature relation and the density of states.
In Section~\ref{sec:grey}, we calculate the greybody factors for particles with different spins and derive the main formula in Eq.~\eqref{eq:gammafac-full}. 
In Section~\ref{sec:evolution}, we choose a few representative model parameters to demonstrate the time evolution of an NEBH. We discuss some phenomenological implications and conclude our paper in Section \ref{sec:conclusion}. In Appendix~\ref{sec:appendix}, we calculate the masses and the corresponding operator scaling dimensions for various fields after dimensional reduction to ${\rm AdS}_2\times S^2$ including a massive gauge boson field.

\section{Properties of charged black holes} \label{sec:property}

Starting with the Einstein-Hilbert action in $(1+3)$-dimension spacetime together with matter field Lagrangian, one has 
\beqa
S = \int d^4x\,\sqrt{-g}\left[ \frac{1}{16\pi\,G}\,R \,+\, \mathcal{L}_{\rm SM} \, +\, \mathcal{L}_{\rm dark} \right] ~. 
\eeqa
Here, the Newton constant is $G = 1/M^2_{\rm pl}$ with $M_{\rm pl} = 1.22 \times 10^{19}$~GeV; $R$ is the Ricci scalar; $g \equiv \mbox{det}(g_{MN})$ with the metric $g_{MN} = (g^{MN})^{-1}$ and the convention of $(-1, +1, +1, +1)$. The Lagrangian $\mathcal{L}_{\rm SM}$ contains all gauge and matter fields in the SM, while the Lagrangian $\mathcal{L}_{\rm dark}$ contains hidden or dark fields beyond the SM. There could be additional non-gravitational interactions between the SM and dark sectors, which are ignored here and not important for the study in this paper. 

We are interested in the properties of a charged or Reissner–Nordstr\"{o}m (RN) black hole with a fixed charge $Q$, which could be the magnetic charge in the SM or other dark gauge charges. Because of the universal properties of a near-extremal charged black hole, the specific type of charges is not important (they do have different phenomenological consequences). The Reissner–Nordstr\"{o}m spacetime metric is
\beqa
ds^2 \, = \, - f(r) \,dt^2 + f(r)^{-1} \, dr^2 + r^2\,(d\theta^2 + \sin^2{\theta} \, d\phi^2) ~,
\eeqa
with 
\beq
f(r) = 1 - \frac{2\,M\,G}{r} + \frac{g^2\,Q^2\,G}{4\pi\,r^2} \, \equiv\, \frac{(r - r_+) (r - r_-)}{r^2} ~.
\eeq
Here, $M$ is the asymptotic mass of the black hole and $Q$ is the black hole charge with $g$ as the charge gauge coupling. The outer and inner horizon radius are given by
\beqa
r_{\pm} = G\,M \pm \sqrt{ G^2 M^2 - \frac{g^2 Q^2\,G}{4\pi} } ~.
\eeqa
The two radii become equal, $r_+ = r_- = r_{\rm e} \equiv G\,M_{\rm e}$, when the black hole mass reaches the extremal limit $M_{\rm e} = g\,Q\,M_{\rm pl}/\sqrt{4\pi}$. Using the ``normal" zeroth law of the black hole dynamics~\cite{Wald:1999vt}, the black hole temperature is 
\beq
T = \frac{r_+ - r_-}{4\pi\,r_+^2}  = \frac{M_{\rm pl}^2}{2\pi}\, \frac{\sqrt{M^2 - M_{\rm e}^2} }{\left( M + \sqrt{M^2 - M^2_{\rm e}} \right)^2} ~. 
\eeq
In the near-extremal limit with $M - M_{\rm e} \ll M_{\rm e}$, the ``semiclassical" relation between $T$ and the extremality parameter or the black hole energy $E \equiv M - M_{\rm e}$ is $E \approx 2\pi^2\,M_{\rm e}^3\,T^2/M^4_{\rm pl}$. In this limit, the Bekenstein-Hawking entropy is $S = \pi\,M_{\rm pl}^2\, r_+^2 = S_0 + 4\pi^2\,M_{\rm e}^3\,T/M_{\rm pl}^4$ with $S_0 = \pi\,M_{\rm e}^2/M_{\rm pl}^2$. 

As pointed out in Refs.~\cite{Preskill:1991tb,Maldacena:1998uz}, the semiclassical description of the charged black hole breaks down when the temperature is below a mass gap scale 
\beqa
\label{eq:lambda-gap}
\Lambda_{\rm gap} \equiv \frac{M_{\rm pl}^4}{M_{\rm e}^3} = \frac{1}{M_{\rm pl}^2\,r_{\rm e}^3} = \frac{(4\pi)^{3/2} \, M_{\rm pl}}{g^3\,Q^3}~. 
\eeqa
Additional quantum effects are important and could change the relations between $E$ and $T$ and between $S$ and $T$. In the near-horizon region with $r - r_{\rm e} \ll r_{\rm e}$ and the low temperature limit with $T \ll 1/r_{\rm e}$, one can define $\rho = r - r_{\rm e}$ and have an approximate metric as
\beqa
\label{eq:metric-near}
ds^2 = - \frac{\rho^2 - 4\pi^2\,r_{\rm e}^4\,T^2}{r^2_{\rm e}} \,dt^2  + \frac{r^2_{\rm e}} {\rho^2 - 4\pi^2\,r_{\rm e}^4\,T^2}\,d\rho^2 + (r_{\rm e} + \rho)^2 \,(d\theta^2 + \sin^2{\theta} \, d\phi^2) ~.
\eeqa
Neglecting the $T^2$ terms and keeping only the leading terms in $\rho$, the above metric can be identified to an AdS$_2 \times S^2$ geometry with both AdS$_2$ and $S^2$ radii as $r_{\rm e}$. 

After a dimensional reduction for the near-extremal case, the action can be matched to the dilaton-gravity model~\cite{Almheiri:2014cka}.  The relevant degree of freedom is the ``Goldstone boson" related to the coset space of time reparametrization symmetry over $SL(2, \mathbb{R})$~\cite{Maldacena:2016upp} or $\mbox{diff}(S^1)/SL(2,  \mathbb{R})$~\cite{Stanford:2017thb}: fixing Poincare coordinates on the hyperbolic disk spontaneously breaks the reparametrization symmetry and the dilation-dependent terms explicitly break this symmetry (see~\cite{Sarosi:2017ykf} for a review). The one-dimensional action for the time reparametrization field is the Schwarzian action, whose path integral can be computed exactly to generate the canonical partition function~\cite{Iliesiu:2020qvm}
\beqa
\label{eq:partition-function}
Z(T) = \left( M_{\rm pl}^2\,r_{\rm e}^3\,T \right)^{3/2} \,e^{S_0 \,-\, M_{\rm e}/T \,+\, 2\pi^2\,M_{\rm pl}^2\,r_{\rm e}^3\,T} ~,
\eeqa
which is valid for $Q \gg 1$ and $T \ll 1/r_{\rm e}$ and is one-loop exact~\cite{Stanford:2017thb}. Here, $S_0$ is the extremal entropy with $S_0 = \pi M_{\rm pl}^2 r_{\rm e}^2$. The entropy and energy as a function of  $T$ are derived as
\beqa
S &=& S_0 \,+\, 4\pi^2\,M_{\rm pl}^2\, r_{\rm e}^3 \, T + \frac{3}{2}\,\ln\left(e\,M_{\rm pl}^2\,r_{\rm e}^3\,T \right) ~, \\
\label{eq:E-T-relation}
E &=& 2\pi^2\, M_{\rm pl}^2\, r_{\rm e}^3 \, T^2 + \frac{3}{2}\,T  = 2\pi^2\,\frac{T^2}{\Lambda_{\rm gap}} \,+\, \frac{3}{2}\,T~. 
\eeqa
The first term of energy comes from the leading semiclassical correction for $T \ll 1/r_{\rm e}$, while the second term linear in $T$ contains the quantum effects and comes from the one-loop backreaction of the dilation and gauge fields on the metric. Note that the second term becomes important when $T \lesssim \Lambda_{\rm gap}$. Also there is no real energy gap as $T$ reaches zero. 

\begin{figure}[t!]
	\centering
	\includegraphics[width=0.7\linewidth]{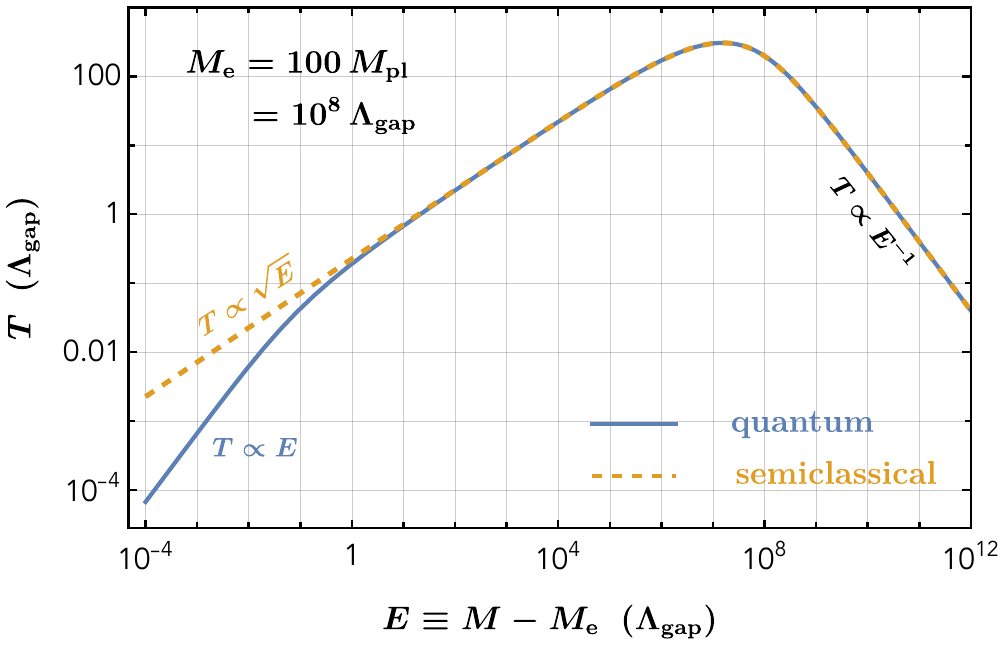}
	\caption{The charged black hole temperature as a function of energy for $M_{\rm e} = 100\,M_{\rm pl}$, which is equivalent to $M_{\rm e} = 10^8 \,\Lambda_{\rm gap}$ and $\Lambda_{\rm gap} = 10^{-6}\,M_{\rm pl}$ from \eqref{eq:lambda-gap}. The power-law behaviors are labelled for the asymptotic regions. In the small energy region the quantum effect turns to reduce the black hole temperature for a fixed energy. 
	 \label{fig:temperature-energy}
	 }
\end{figure}

In Fig.~\ref{fig:temperature-energy}, we show the charged black hole temperature as a function of energy $E$ for the extremal mass $M_{\rm e} = 100\,M_{\rm pl}$ or $Q = \mathcal{O}(100)$ for $g = \mathcal{O}(1)$. Here, we have shown both relations from the quantum corrected one and the semiclassical one. For a larger energy or the black hole mass farther away from the extremal mass, the semiclassical description works very well. This is not true any more when the black hole energy is below around the black hole mass gap scale $\Lambda_{\rm gap}$, where the quantum effect becomes important and the scaling of temperature in energy changes from $T \propto \sqrt{E}$ (semiclassical) to $T \propto E$ (quantum). As a result and for a fixed energy or mass, the black hole has a lower temperature (and likely less Hawking radiation) compared to the semiclassical result. 

Using the relation of $Z(T) = \int_0^\infty dE \, \rho(E)\, e^{-(E + M_{\rm e})/T}$ and Eq.~\eqref{eq:partition-function}, the density of states for a fixed charge $Q$ is 
\beqa
\label{eq:density-of-states}
\rho(E) = e^{S_0} \,\frac{M_{\rm pl}^2\, r_{\rm e}^3}{\sqrt{2}\,\pi^{3/2}} \, \sinh{\left(2\pi\sqrt{2\,M_{\rm pl}^2 \,r_{\rm e}^3 \,E}\right)}  \,=\,  \frac{e^{S_0} }{\sqrt{2}\,\pi^{3/2}\,\Lambda_{\rm gap}} \, \sinh{\left(2\pi\sqrt{2\,E/\Lambda_{\rm gap}}\right)} ~, 
\eeqa
which is valid for $T \ll 1/r_{\rm e}$. In the low-energy limit with $E \sim T \ll \Lambda_{\rm gap}$, the energy of state has the scaling of $\rho(E) \propto \sqrt{E}$ and smoothly goes to zero. This density of state function may suggest that there is a continuous spectrum, which is a misleading interpretation. The actual quantum mechanical system or the ultraviolet theory could have a discrete spectrum and lead to \eqref{eq:partition-function} and \eqref{eq:density-of-states} as a limit of other variables~\cite{Stanford:2017thb}. One condensed-matter system leads to the similar spectrum is the SYK model~\cite{Sachdev:1992fk,Kitaev} (see Ref.~\cite{Chowdhury:2021qpy} for a review).

\section{Greybody factors for particles with different spins}\label{sec:grey}
Given the new understanding of near-extremal charged black holes at the quantum level, their Hawking radiation of massless or light particles can be calculated in the dual CFT theory. The relevant quantity is the greybody factor which characterizes the deviation of the black hole emitted spectrum from a blackbody one due to the propagation of particles from the event horizon to the far infinity region. For a non-rotating charged black hole and radiating uncharged quanta, the greybody factor is defined using the expected number of particles in radiation as~\cite{Page:1976df} 
\beqa
\label{eq:expected_N}
\langle N_{s l p }(\omega) \rangle = \frac{\Gamma_{s l p}(\omega, T, r_{\rm e})}{2\pi} \, \frac{1}{e^{\omega/T}  \mp 1} ~,
\eeqa
where $s$ denotes the spin of the radiated species; $l\ge s$ with $l -s \in \mathbb{Z}$  is the spherical harmonic mode or the total angular moment; $p$ is the specific polarization or helicity of the radiated particle; ``$+$" in the denominator is for fermion and ``$-$" is for boson. The greybody factor $\Gamma$ is a function of radiation energy $\omega$ and the black hole mass $M$ and charge $Q$, which we represent using $T$ and $r_{\rm e}$.

On the gravitational side, one can obtain the semiclassical results by solving the (Teukolsky~\cite{Teukolsky:1973ha}) equations of motion of the matter fields. One first solves the field radial functions in the near-horizon region and the far region, and then match the two solutions in the transition region to obtain the absorption cross section $\sigma^{\rm abs}$, which is related to the greybody factor as $\sigma^{\rm abs}_{sp}(\omega, T, r_{\rm e}) = \pi \,\omega^{-2}\, \sum_l  (2 l  + 1)\Gamma_{s l p}(\omega, T, r_{\rm e})$. Following this procedure, Ref.~\cite{Page:1976df} obtained a simple analytic formula for the greybody factor in the limit of $4\pi r_+^2 \omega T \ll 1$ or $\omega, T \ll 1/r_{\rm e}$ for a near-extremal charged black hole. After some algebras, we rewrite the formula in Ref.~\cite{Page:1976df} as (the greybody factors for a higher value of $\omega$ have numerical solutions and can be found in Ref.~\cite{Arbey:2019mbc})
\beqa\label{eq:gammafac-semi}
\Gamma_{s l p}^{\rm semi} &=& \frac{1}{\pi}\,\left[\frac{\Gamma(l-s+1)\,\Gamma(l+s+1)}{\Gamma(2l+1)\,\Gamma(2l+2)}\right]^2\,\nonumber \\
&&\hspace{1cm}\times\,\left(8\pi r_+^2\, \omega\,T\right)^{2l+1} \,\Big{|}\Gamma\Big{(}l+ 1 + \frac{i\omega}{2\pi T}\Big{)}\Big{|}^{2}\, \times \,\begin{cases}\sinh\big{(}\frac{\omega}{2 T}\big{)} & \text{(boson)} \\ \cosh\big{(}\frac{\omega}{2 T}\big{)} & \text{(fermion)} \end{cases} ~.
\eeqa
In the limit of $\omega \ll 4\pi T$, the above formula can be further simplified. For the dominant contribution with $l = s$, we list the greybody factors for different spins in Table~\ref{tab:semiclassical} (see also Ref.~\cite{Cvetic:1997ap}). 

%
\begin{table}[hb!]
	\centering
	\renewcommand{\arraystretch}{2.0}
		\begin{tabular}{ c | c | c | c | c  }
			\hline \hline
spin ($s =$)   &   0 & $\frac{1}{2}$ & 1 & 2 
  \\ \hline
$\Gamma_{ssp}^{\rm semi}$ & $4\,r_+^2\,\omega^2$ & $\frac{1}{4}\,(r_+ - r_-)^2 \,\omega^2$ & $\frac{4}{9}\,r_+^2\,(r_+ - r_-)^2\,\omega^4$ & $\frac{4}{225}\,r_+^2(r_+ - r_-)^4\,\omega^6$  \\
  \hline  \hline 
		\end{tabular}
	\caption{The greybody factors for various massless spin particles with $l = s$ in the limit of $T \ll 1/r_{\rm e}$ and $\omega \ll 4\pi T$ and based on the semiclassical calculation, which is valid for $T \gg \Lambda_{\rm gap}$. 
		} \label{tab:semiclassical}
\end{table}
%

The absorption cross section $\sigma^{\rm abs}$, or the greybody factor $\Gamma_{s l p}$ can also be derived via $\text{AdS}_{d+1}/\text{CFT}_{d}$ correspondence (or  $\text{AdS}_{2}/\text{CFT}_{1}$ for the near-extremal charged black hole case). The basic logic is that the near horizon region of the black hole $\text{AdS}_{d+1}$ is dual to a conformal field theory $\text{CFT}_{d}$ on the boundary of the $\text{AdS}_{d+1}$ space~\cite{Maldacena:1997re,Aharony:1999ti}. The absorption of the probe particles by the black hole then can be described by the interaction $S_{\rm int} = \int dt dz\, \phi(t, z=0)\,\mathcal{O}(t)$, with $\mathcal{O}$ denoting the dual CFT operator with a scaling dimension of $\Delta_{\mathcal{O}}$ sitting at the boundary ``brane" with $z=0$. For a radiated quantum with an energy $\omega$ and using the Fermi's Golden Rule, the transition amplitude from an initial state $|i\rangle$ to a final state $|f\rangle$ is $\mathcal{M}_{i \rightarrow f} \sim \langle f| \int dt \,\mathcal{O}(t) e^{-i\omega t} |i \rangle$. After squaring the matrix element, summing over the final state and averaging over the initial states with thermal ensemble, the total absorption probability is~\cite{Maldacena:1997ih,Gubser:1997cm}
\beqa
\label{eq:gammacft}
\Gamma^{\rm abs} &\propto& \int dt\,e^{-i \omega t} \sum_i e^{-E_i/T} \langle i| \mathcal{O}(t)\mathcal{O}(0) | i \rangle \propto \omega^{2\Delta_{\mathcal{O}}-1} \mathcal{F}_{\Delta_{\mathcal{O}}}(T,\omega,r_{\rm e}) \nonumber \\
&&\hspace{1cm}\mbox{with}~~\mathcal{F}_{\Delta_{\mathcal{O}}}(T,\omega,r_{\rm e}) \equiv \int dt\,e^{-i\omega t}\big{[}G^{+}_{\Delta_{\mathcal{O}}}(T, t)-
G^{-}_{\Delta_{\mathcal{O}}}(T, t)\big{]}  ~.
\eeqa
Here, the overall constant factor depends on $s$ and $l$ and is not important for understanding the $T$ and $\omega$ dependence of the greybody factor. $G^{\pm}_{\Delta_{\mathcal{O}}}(T, t)$ is a thermal two-point function with a scaling dimension $\Delta_{\mathcal{O}}$. Based on the Schwarzian action, equivalent to the CFT$_1$ dual to AdS$_2$, the general expression for this thermal two-point function is calculated at loop level in Ref.~\cite{Mertens:2017mtv,Kitaev:2018wpr} and is given by
\beqa
G^{\pm}_{\Delta_{\mathcal{O}}}(T, t) &=& \frac{1}{2\sqrt{\pi}}\left(\frac{\Lambda_{\rm gap}}{2\pi^{2} T}\right)^{3/2} \, e^{-\frac{2\pi^{2}T}{\Lambda_{\rm gap}}}\,\int dq_{1} dq_{2}\,\sinh(2\pi \sqrt{q_{1}})\,\sinh(2\pi \sqrt{q_{2}})\,e^{\pm it (q_{1}-q_{2})\,\Lambda_{\rm gap} -\frac{ q_{2}\,\Lambda_{\rm gap}}{2\,T}}   \nonumber \\
&&\hspace{+4.5cm}\times \, \frac{1}{\Gamma(2\,\Delta_{\mathcal{O}})}\,\prod_{\varkappa=\pm, \varsigma=\pm}\Gamma[\Delta_{\mathcal{O}} + \varkappa\, i(\sqrt{q_{1}} + \varsigma\, \sqrt{q_{2}})] ~,
\eeqa
which is valid in the limit of $T \ll 1/r_{\rm e}$ or the near-extremal region. 
Substituting $G^{\pm}_{\Delta_{\mathcal{O}}}(T, t)$ into \eqref{eq:gammacft} and first performing an integration in $t$,  the function $\mathcal{F}$ is calculated to be
\beqa
\label{eq:F-function}
\mathcal{F}_{\Delta_{\mathcal{O}}}(T,\omega,\Lambda_{\rm gap}) &=& \frac{\sqrt{\pi}}{\Lambda_{\rm gap}}\, \left(\frac{\Lambda_{\rm gap}}{2\pi^{2} T}\right)^{3/2} \, e^{-\frac{2\pi^{2}T}{\Lambda_{\rm gap}}}\, \frac{(e^{\frac{\omega }{2 T}}  - 1 )}{\Gamma(2\,\Delta_{\mathcal{O}})}\, \times  \\
&&\hspace{-3.8cm}\int^\infty_{\omega/\Lambda_{\rm gap}} dq \, \sinh(2\pi\sqrt{q}) \, \sinh\left(2\pi\sqrt{q- \frac{\omega}{\Lambda_{\rm gap} } }\right) \, e^{-\frac{ q\,\Lambda_{\rm gap}}{2\,T}}\,\prod_{\varkappa=\pm, \varsigma=\pm}
\Gamma\left[\Delta_{\mathcal{O}} + \varkappa\, i\left(\sqrt{q} + \varsigma\, \sqrt{q - \frac{\omega}{\Lambda_{\rm gap} } } \right)\right]~, \nonumber
\eeqa
which is again valid in the limit of $T \ll 1/r_{\rm e}$.
In the high temperature limit with $T \gg \Lambda_{\rm gap}$, the temperature and frequency dependence for $\Gamma^{\rm abs}$ matches the dependence of the semi-classical results in \eqref{eq:gammafac-semi}. Using the ratio of $\mathcal{F}_{\Delta_{\mathcal{O}}}(T,\omega,\Lambda_{\rm gap})$ over its high-temperature behavior, the greybody factor including quantum corrections is 
\beqa \label{eq:gammafac-full}
\Gamma_{slp}^{\rm full} = \frac{\mathcal{F}_{\Delta_{\mathcal{O}}}(T,\omega,\Lambda_{\rm gap}) }{\mathcal{F}_{\Delta_{\mathcal{O}}}(T\gg \Lambda_{\rm gap},\omega,\Lambda_{\rm gap}) } \, \times  \Gamma_{slp}^{\rm semi} ~.
\eeqa
The above equation \eqref{eq:gammafac-full} together with  \eqref{eq:gammafac-semi} and \eqref{eq:F-function} provide the full results of the greybody factors in the limit of $T \ll 1/r_{\rm e}$. Note that for $\Delta_{\mathcal{O}} = 1$ case, the denominator has a simple formula $\mathcal{F}_{\Delta_{\mathcal{O}} = 1}(T\gg \Lambda_{\rm gap},\omega,\Lambda_{\rm gap}) = 2\pi\,\Lambda_{\rm gap}^{-2}\,\omega$. For a general $\Delta_{\mathcal{O}}$, one could use the method of steepest descent or the Laplace's method to perform the integration in $q$ and have 
\begin{align}
\mathcal{F}_{\Delta_{\mathcal{O}}}(T\gg \Lambda_{\rm gap},\omega,\Lambda_{\rm gap}) \approx (e^{\frac{\omega}{2T}} -1)\times \,\begin{cases}
2\pi^{3}\,T^{2}\,\Lambda_{\rm gap}^{-3}~, & \Delta_{\mathcal{O}} = 3/2 ~,\\
\frac{32}{3} \pi^{3}\,T^{3}\,\Lambda_{\rm gap}^{-4}~, & \Delta_{\mathcal{O}} = 2 ~,\\
\frac{512}{15} \pi^{5}\,T^{5}\,\Lambda_{\rm gap}^{-6}~, & \Delta_{\mathcal{O}} = 3~.\\
\end{cases}
\end{align}

For particles with different spins, the scaling dimension $\Delta_{\mathcal{O}}$ for its corresponding operator can be derived using the routine procedure of AdS/CFT. One adds the matter field in the $1+3$ theory and performs a dimensional reduction to AdS$_2 \times S^2$. For bosons, the scaling dimension is given by $\Delta_{\mathcal{O}} = (1 + \sqrt{1 + 4\,m^2 \,r^2_{\rm e}})$ with $m^2$ as the particle mass in AdS$_2$ spacetime. For fermions, $\Delta_{\mathcal{O}} = (1 + 2\, |m|\,r_{\rm e})/2$. We perform the dimensional reduction in Appendix \ref{sec:appendix} for spin-$0, 1/2, 1, 2$ massless particles and massive spin-1 bosons and find a universal relation between $\Delta_{\mathcal{O}}$ and $s, l$ as
\beqa
\Delta_{\mathcal{O}}^{sl} = l + 1 \,, \qquad \mbox{for}\,~\,l \ge s \,\, \mbox{and}\,\, l - s \in \mathbb{Z} ~. 
\eeqa
For the massive spin-1 particles including $W^\pm$ and $Z$ bosons, they contain the Goldstone boson (GB), which can be effectively treated as a spin-0 state with the lowest scaling dimension of $\Delta_{\mathcal{O}} = 1$ in the limit of $1/r_{\rm e} \gg M_W, M_Z$ (see Appendix~\ref{sec:spin-1-massive} for detail). 

\begin{figure}[t!]
	\includegraphics[width=0.7\textwidth]{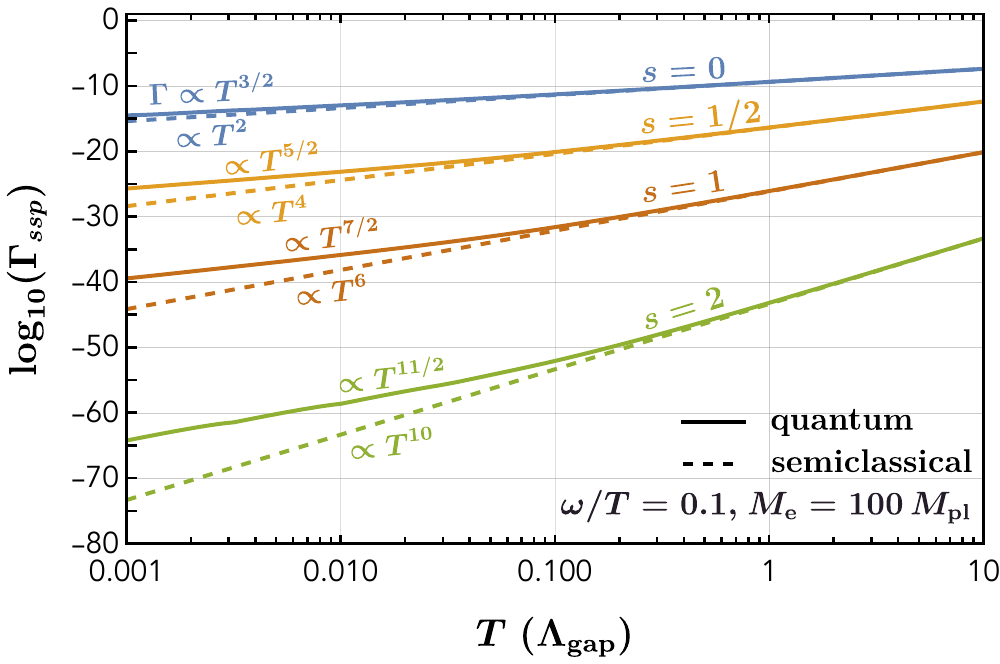}
	\caption{Greybody factors $\Gamma_{ssp}$ as function of $T$ for massless particles with different spins for $M_{\rm e}=100\,M_{\rm pl}$ and $\omega/T=0.1$. The quantum corrected ones from \eqref{eq:gammafac-full} and with a scaling of $\Gamma_{ssp}\propto T^{2s+3/2}$  in the limit of $T\ll \Lambda_{\rm gap}$ are shown in solid lines, while the semiclassical results from \eqref{eq:gammafac-semi} and with a scaling of $\Gamma_{ssp}\propto T^{4s+2}$ are shown in dashed lines. }
	\label{fig:greybodycomp}
\end{figure}

In Fig.~\ref{fig:greybodycomp}, we show a comparison between the full result with quantum corrections and the semiclassical one for $l = s$ case and for different spins and a fixed ratio of $\omega/T = 0.1$. The agreement is very well for $T \gtrsim \Lambda_{\rm gap}$, while they differ a lot in the low-temperature region with $T \ll \Lambda_{\rm gap}$. For $l = s$, $T \ll \Lambda_{\rm gap}$ and a fixed ratio of $\omega/T$, one has $\Gamma_{ssp}^{\rm full}  \propto r_{\rm e}^{4s + 2}\,\Lambda_{\rm gap}^{2s +1/2}\,T^{2 s + 3/2}$ instead of the semiclassical scaling of $\Gamma_{ssp}^{\rm semi}  \propto (r_{\rm e}\,T)^{4 s + 2}$. As a result, the greybody factors with the quantum corrections are much larger than the semiclassical one for $T \ll \Lambda_{\rm gap}$ or the late stage of charged black hole evolution. Also note that the greybody factor for radiating a spin-0 particle is much larger than the one for radiating a higher-spin particle. 
For instance, the ratio of greybody factors for $s=1/2$ over $s=0$ is $\sim r^2_{\rm e} \,\Lambda_{\rm gap}\,T = T/(Q\,M_{\rm pl}) < 1/Q^4$ for $T < \Lambda_{\rm gap}$ and is highly suppressed for $Q \gg 1$. Given that the SM contains one elementary scalar particle, the Higgs boson, radiating the Higgs boson (also the scalar degrees contained in the Higgs doublet) could be the dominant channel for the a near-extremal charged black hole to reduce its energy when its temperature is much above the electroweak scale. When the black hole energy is below the Higgs and weak gauge boson masses, those radiation channels are kinematically forbidden and the radiation channels into other SM lighter states become important. 

\section{Evolution of near-extremal charged black holes}\label{sec:evolution}

The time evolution of the near-extremal black hole energy is related to the expected number of particles $\langle N_{slp}(\omega)\rangle$ in \eqref{eq:expected_N} and is given by
\beqa
\label{eq:evolution}
\frac{dE}{dt} = -\sum_{i} \sum_{l} \int_{0}^{E}\,d\omega\, \omega\, g_{i}\,\langle N^{i}_{slp}(\omega)\rangle ~,
\eeqa
where $i$ runs over SM or dark species with a degree of freedom $g_{i}$. In the relativistic limit, one has $g_i = 1$ for a real scalar, $2$ for a Weyl fermion, $2$ for a massless gauge boson, and $2$ for the graviton. Note that the integration upper limit is $E$, which can be approximately replaced by $\infty$ in the semiclassical limit with $E \gg T$, since the quanta with $\omega \gg T$ is Boltzmann suppressed.  In the near-extremal region with $T \ll 1/{r_{\rm e}}$ and because integration in \eqref{eq:evolution} is dominated in the region of $\omega \sim T$, the sum  $\sum_{l}$ is governed by the smallest one with $l = s$.  The black hole mass or energy $E$ as a function of $t$ can be estimated analytically to have a power-law behavior. We also note that when $T \lesssim \Lambda_{\rm gap}$, the black hole energy changes significantly after each emission and its evolution becomes stochastic. A discretized implementation of Eq.~\eqref{eq:evolution} will be adopted to understand the later stage of black hole evolution.

For a fixed $Q$ and starting with a near-extremal black hole with $E_{0} = M_0 - M_{\rm e}$ much above $\Lambda_{\rm gap}$ at the initial time $t_0 = 0$, the evolution of black hole first follows the semiclassical one. For the case with radiation dominated by one particle with a spin $s$ and with the temperature much above the particle mass, the energy as a function of time is 
\beqa
E^{\rm semi}(t) = \left( E_0^{-(2s+1)} + c_1\,r_{\rm e}^{4s+2}\,\Lambda_{\rm gap}^{2s + 2}\,t \right)^{-1/(2s + 1)} ~,
\eeqa
with $c_1$ dependent on $s$ and other numerical numbers. For $s=0$, one has $c_{1}= g_{i}/(30\pi)$.
 After $t \gtrsim t_1 \equiv c_1^{-1}\,(r_{\rm e}^2 \, \Lambda_{\rm gap}\,E_0)^{-(2s+1)}\,\Lambda_{\rm gap}^{-1}$, the $E^{\rm semi}(t)$ approaches a simple power-law behavior in $t$ as $E^{\rm semi}(t)  \propto r_{\rm e}^{-2}\,\Lambda_{\rm gap}^{-1}\,(\Lambda_{\rm gap}\,t)^{-1/(2s+1)}$. As the energy drops and becomes comparable to the mass gap scale $\Lambda_{\rm gap}$, the quantum effects become important, which happens at a time $t_2 \equiv (\Lambda_{\rm gap}\,r_{\rm e})^{-(4s+2)}\,\Lambda_{\rm gap}^{-1}\,c_1^{-1}$. For $t \gtrsim t_2$, the energy as a function of time is 
\beqa
\label{eq:E-in-t-quantum}
E^{\rm quantum}(t) = c_2 \, r_{\rm e}^{-2}\,\Lambda_{\rm gap}^{-1}\, \left( r_{\rm e}^{-3}\,\Lambda_{\rm gap}^{-2}\,t \right)^{-1
/(2s + 5/2)} ~,
\eeqa
with the order-one number $c_2$ dependent on the spin $s$ and other numerical numbers. For $t_2 < t_{\rm univ.} = 13.7\,\mbox{Gyr}$, the black hole in the current universe could already reach the quantum-dominant region. Otherwise, its evolution is still in the semiclassical region. Converting to the extremal mass, one has 
\beqa
M_{\rm e} < c_1^{1/(8s + 7)} \,\left(M_{\rm pl} \, t_{\rm univ}\right)^{1/(8s + 7)}\,M_{\rm pl} ~,
\eeqa
to have the black hole in the quantum-effect dominated region in the current universe. Note that $M_{\rm pl} \, t_{\rm univ} \approx 8\times 10^{60}$. Since the greybody factor is larger for a smaller-spin particle, the evolution is dominated by radiating the SM Higgs boson and the ``scalar" degree of freedom in massive gauge bosons. For $s=0$, we have the condition to be $M_{\rm e} < c_1^{1/7} \times\,5 \times 10^8\,M_{\rm pl}$.

After $t_2$ or the black hole energy drops below $\Lambda_{\rm gap}$, the black hole evolution becomes stochastic. This is because its energy $E \approx 3\,T/2$ [see \eqref{eq:E-T-relation}] when the quantum effects dominate. The radiated quantum has a frequency of $\omega \sim T$ or comparable to the temperature. So, the black hole energy and temperature changes by an order-one factor after emitting each quantum. The charged black hole evolution follows a discretized and stochastic way: different charged black holes evolve in their own ways. Because the greybody factor is much larger for radiating a particle with the smallest scaling dimension $\Delta_{\mathcal{O}}$, the black hole evolution is dominated by radiating the Higgs boson and one degree of freedom in $W^\pm$ and $Z$ bosons (see Appendix~\ref{sec:spin-1-massive} for detailed discussion), which have nonzero masses $m_h =125$~GeV, $M_{W}=80.4$~GeV and $M_Z=90.2$~GeV. When the black hole energy drops below $M_W$, the Hawking radiation of the Higgs boson and massive gauge bosons is kinematically forbidden. The charged black holes are ``stuck" in an approximately steady state with a fixed energy and temperature below $M_W$. Different black holes reach different steady energies. Radiation of SM fermions can potentially keep decreasing the black hole energy, but it will only happen at a much later time beyond the age of the universe [one can check this using Eq.~\eqref{eq:E-in-t-quantum} with $s=1/2$]. 

\begin{figure}[th!]
	\includegraphics[width=0.7\textwidth]{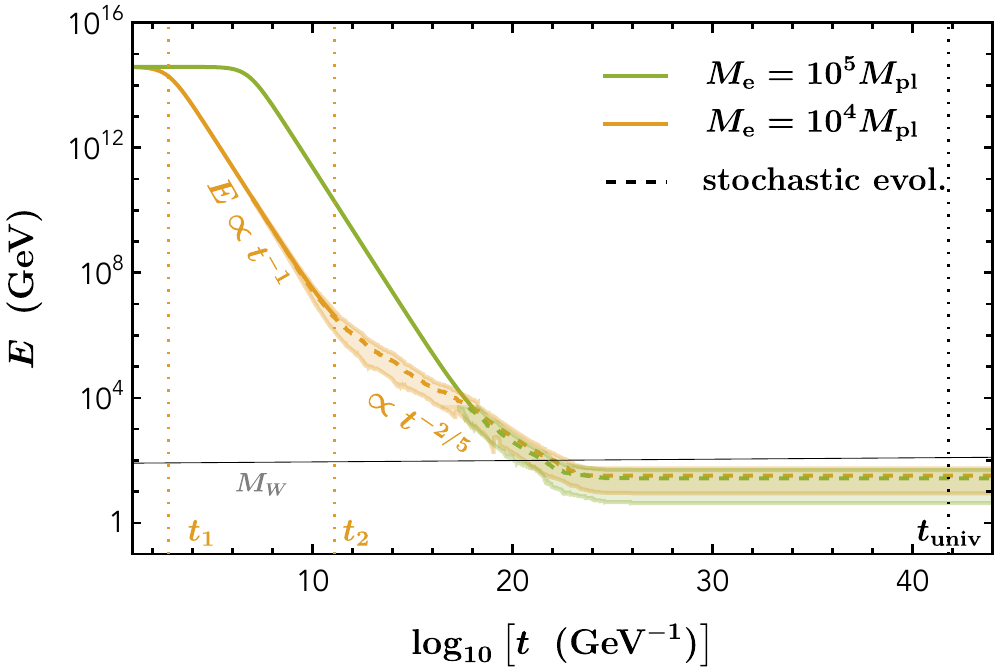}
	\caption{Time evolutions of charged black holes for two different charges or extremal masses $M_{\rm e} = 10^5\,M_{\rm pl}$ and $10^4\,M_{\rm pl}$. The initial energy is chosen to be $4\times 10^{14}$\,GeV. The Hawking radiation is dominated by the Higgs boson and one degree of freedom in $W^\pm$ and $Z$ gauge bosons. SM fermions, photon and graviton play negligible roles. After $t_1$, the black hole starts to decrease energy dramatically and follows a power law $E \propto t^{-1}$ governed by the semiclassical result. After $t_2$ when $E \approx \Lambda_{\rm gap}$, the quantum effects become important. The evolutions become stochastic with the central values shown in the dashed line  ($E \propto t^{-2/5}$ just after $t_2$)  and one-sigma variance band shown in the shaded region. After black hole energy drops below the $W$ gauge boson mass $M_W = 80.4$~GeV, its evolution is approximately stopped. $t_{\rm univ}$ denotes the age of current universe.
	}
	\label{fig:evolution}
\end{figure}

In Fig.~\ref{fig:evolution}, we show the evolutions of charged black holes for two different extremal masses $M_{\rm e} = 10^5\,M_{\rm pl}$ and $10^4\,M_{\rm pl}$, which have $\Lambda_{\rm gap} = 1.22 \times 10^4\,\mbox{GeV}$ and $1.22 \times 10^7\,\mbox{GeV}$, respectively. A large number of black holes have been simulated and have the averaged energy shown in the dashed line and the one-sigma variance shown in the shaded band. For an individual black hole, its evolution has a multi-step-like behavior in time. Its evolution stops after its energy drops below the $W$ gauge-boson mass $M_W$ shown in the horizontal gray line (until a much later time beyond the age of the universe it starts to radiate SM fermions). For the two extremal masses, the final black hole energies are $26 \pm 22$~GeV for $M_{\rm e} = 10^5\,M_{\rm pl}$ and $32 \pm 22$~GeV for $M_{\rm e} = 10^4\,M_{\rm pl}$. 

The charged black holes at the current universe can radiate other SM particles with the SM fermions as the leading channels. Since $T\ll \Lambda_{\rm gap}$, the quantum effects control the relation of $E \approx 3\,T/2$ and the greybody factors. The emission rate to fermions is calculated to be
\beqa
\mathcal{R}_{1/2} &\equiv& g\, \int_{0}^{E}\,d\omega\, \,\langle N_{\frac{1}{2}\frac{1}{2}p}(\omega)\rangle \approx 0.0046\,g\, r_{\rm e}^4\,\Lambda_{\rm gap}^{3/2}\, E^{7/2}  = 0.0046\,g\, M_{\rm e}^{-1/2}\,M_{\rm pl}^{-2}\, E^{7/2}  \nonumber \\
&=&  1.0 \times 10^{-19}\,\mbox{s}^{-1}\, \times \left( \frac{g}{78}\right) \, \left( \frac{10^4\,M_{\rm pl}}{M_{\rm e}} \right)^{1/2} \, \left( \frac{E}{100\,\mbox{GeV}} \right)^{7/2} ~,
\label{eq:emitting-rate-fermion}
\eeqa
where $g = 78$ includes all SM fermion degrees of freedom except the top quark, which is heavier than the black hole energy $E \lesssim M_W = 80.4$~GeV. Inverse of this rate is longer than the age of the universe, which justifies our previous ignorance of radiation to fermions. 

\section{Discussion and conclusions} \label{sec:conclusion}

Our discussion about the charged black hole evolution has ignored the potential Schwinger effects~\cite{Schwinger:1951nm} to discharge the black hole. This could be justified for some dark or hidden charged black holes like the one with a dark QED and a very heavy dark electron~\cite{Bai:2019zcd}. Alternatively, the black hole may also carry SM magnetic~\cite{Maldacena:2020skw,Bai:2020ezy} or hidden magnetic charge with the magnetic monopole mass much above the black hole magnetic field at the event horizon. Note that those charged black holes, if produced in the early universe, could become an NEBH after Hawking radiation and behave as a primordial extremal black hole, a natural dark matter candidate~\cite{Bai:2019zcd}. The greybody factors and the subsequent evolution for black holes calculated in this paper are therefore important for us to search for those primordial charged black holes. Although the emission rate of NEBHs in the current universe from Eq.~\eqref{eq:emitting-rate-fermion} is too small for an indirect detection of NEBHs from Hawking radiated photons and neutrinos, a binary and merging system of NEBHs with opposite-sign charges can have a violent and transient event that could be searched for via a multi-messenger method~\cite{Bai:2019zcd,AyalaSolares:2019iiy}. If such an event is discovered, the temporal information of the spectrum could be used to identify both the semiclassical and quantum-dominated regions of charged black hole Hawking radiation. 

In summary, we have calculated the greybody factors for a near-extremal charged black hole radiating fields with different spins and including the quantum effects. For the black hole temperature below the mass gap scale, both the energy-temperature relation and greybody factors are modified with respect to the semiclassical results. As a result, the black hole evolution is ``slowing down": it takes a longer time to reach the extremal state. Assuming SM particles as the main radiation channels, we have found that the SM Higgs and massive bosons are most important for an NEBH with $M_W < T < \Lambda_{\rm gap}$ and when the quantum effects become important. For a wide range of black hole extremal masses, the charged black hole in the current universe is ``stuck" to a near-extremal state with a temperature below the $W$-gauge boson mass and around 30 GeV, but with a large variance (around 20 GeV) after sampling a large number of black holes. For an individual black hole, its evolution shows a stochastic and step-like way to Hawking radiate quanta and reduce its energy.

\vspace{1cm}
\subsubsection*{Acknowledgments}
We thank Nicholas Orlofsky for useful discussion. The work  is supported by the U.S.~Department of Energy under the contract DE-SC-0017647.  
 
\appendix
\section{Dimensional reduction to ${\rm AdS}_2\times S^2$}\label{sec:appendix}
In this section we calculate the scaling dimensions of operators dual to the probe fields of various spins used to calculate the greybody factors in Section~\ref{sec:grey}. We start with massless probe fields in ($1$+$3$)-dimension black hole spacetime and perform the Kaluza-Klein (KK) dimensional reduction over an  $S^{2}$ sphere to obtain the tower of KK states and their masses in $\text{AdS}_{2}$. We pay special attention to physical and propagating states in the bulk of $\text{AdS}_{2}$. Some early and related literature can be found in Refs.~\cite{Michelson:1999kn,Larsen:2014bqa}. To understand black hole radiation of massive gauge bosons and the Higgs boson, we use a spontaneously breaking $U(1)$ gauge theory as an example. 
After we obtain the physical state masses in $\text{AdS}_{2}$, we use the standard holographic dictionary~\cite{Aharony:1999ti} to obtain the scaling dimensions of operators in CFT$_1$.

\subsection{Spin 0}
\label{sec:spin-0}
The metric of $\text{AdS}_{2}\times S^{2}$ is parametrized as
\begin{equation}
\label{eq:metric-1}
ds^{2}\equiv g_{4M N}dX^{M}dX^{N} = g_{2\mu\nu}(x) dx^{\mu}dx^{\nu}+ r_{\rm e}^{2}\,h_{pq}(y)dy^{p}dy^{q} ~,
\end{equation}
where $X=(x,y)$ with $x=(t,\rho)$ and $y=(\theta,\phi)$ as the $\text{AdS}_{2}$ and $S^{2}$ coordinates, respectively. The indexes  for $\text{AdS}_{2}$ is $\mu, \nu$; the indexes for $S^2$ are $p, q$; the indexes for the 4-dimensional coordinate are $M, N$ with $M=(\mu, p)$.  The metric for $\text{AdS}_{2}$ is $g_{2\mu\nu}=\mbox{diag}(-\rho^{2}/r_{\rm e}^{2}, r_{\rm e}^{2}/\rho^{2})$ and the metric for $S^{2}$ is $r^2_{\rm e}\,h_{pq}=r^2_{\rm e}\,\mbox{diag}(1,\sin^2{\theta})$, which matches the near-horizon metric of the near-extremal black hole in Eq.~\eqref{eq:metric-near}. 

For later calculation convenience, we write down the nonzero Christoffel symbols for both $\text{AdS}_{2}$ and  $S^{2}$: $\Gamma^t_{t\rho} = \Gamma^t_{\rho t} = \rho^{-1}$,  $\Gamma^\rho_{tt} = \rho^3/r_{\rm e}^4$ and $\Gamma^\rho_{\rho\rho}=-\rho^{-1}$; $\Gamma^{\theta}_{\phi\phi} = - \sin{\theta} \cos{\theta}$ and $\Gamma^{\phi}_{\theta\phi} = \Gamma^{\phi}_{\phi\theta} = \cot{\theta}$. The Ricci tensor and scalar for $\text{AdS}_{2}$ are $R_{\mu\nu} = \mbox{diag}(\rho^2/r_{\rm e}^4, - \rho^{-2})$ and $R = - 2/r_{\rm e}^2$, while for $S^{2}$ one has $R_{pq} = \mbox{diag}(1, \sin^2\theta)$ and $R = 2/r^2_{\rm e}$.

For a 4-dimensional real massless scalar and using the spherical harmonics function as the basis function in $(\theta, \phi)$, one has 
\beqa
\Phi(x, y) = \phi_{l m}(x) \, Y^m_l(y)~,
\eeqa
with the Einstein summation and  $\phi_{l-m} = \phi_{lm}^\dagger$. Starting with 4-dimensional action and performing the solid angle integration, one has 
\beqa
S &=& \int d^2x d^2y \,\sqrt{-g_4} \, \frac{1}{2}\,g_4^{MN} \partial_M \Phi \partial_N \Phi  
~, \nonumber \\
&=& \int d^2x \sqrt{-g_2} \,r_{\rm e}^2 \left( \frac{1}{2}\,g_2^{\mu\nu} \partial_\mu \phi^\dagger_{lm} \partial_\nu \phi_{lm} \,+ \, \frac{1}{2}\, \frac{l(l+1)}{r_{\rm e}^2}\, \phi^\dagger_{l m} \phi_{lm}   \right) ~,
\eeqa
with $\sqrt{-g_4} \equiv \sqrt{-\mbox{det}(g_4)} = r_{\rm e}^2\,\sin{\theta}$. So, the masses of the KK modes are $m^2 = l (l + 1)/r_{\rm e}^2$. Using the standard holographic dictionary, the scaling dimensions of corresponding operators are given by $\Delta_{\mathcal{O}} = \frac{1}{2}(d+\sqrt{d^{2}+4\,m^{2}\,r_{\rm e}^{2}})$ with $d=1$ for AdS$_2$~\cite{Aharony:1999ti}. So, for the spin-0 field, the scaling dimensions are $\Delta_{\mathcal{O}} = l + 1$ with the lowest dimension as $\Delta_{\mathcal{O}} = 1$ for $l = 0$.

\subsection{Spin 1/2}
For the spin-1/2 fermion case, we follow the notation in \cite{Maldacena:2018gjk,Bai:2020spd}. The 4-dimensional action is 
\begin{equation}
S = \int d^{4}X \sqrt{-g_{4}}\,\overline{\Psi} \,i\gamma^{A}\,{\rm e}^{M}_{A}\,D_{M}\,\Psi ~. 
\end{equation}
Here, we use $M, N$ to denote curved space indices and $A, B$ for the flat space indices. $\gamma^{A}$ is the gamma matrix in the flat space with $\Sigma_{AB} = \frac{i}{2}[\gamma_A, \gamma_B]$.  With the conversion in \cite{Freedman:2012zz}, they are given by
\begin{align}
\gamma^{0}&=i\, \sigma_{x}\otimes \mathbb{I}_{2}\,, \quad \gamma^{1}=\sigma_{y}\otimes \mathbb{I}_{2}\,, \quad \gamma^{2}=\sigma_{z}\otimes\sigma_{x}\,, \quad \gamma^{3}=\sigma_{z}\otimes\sigma_{y}~, \\
\Sigma_{01} &= -i\, \sigma_{z}\otimes \mathbb{I}_{2}\,, \quad \Sigma_{23}=- \mathbb{I}_{2}\otimes \sigma_{z} ~,
\end{align}
where $\sigma_{x, y, z}$  are the Pauli matrices and $\mathbb{I}_{2}$ is the $2\times2$ identity matrix. For convenience, we change the metric notation as 
\beqa
ds^2 = e^{2\,\sigma(q)}\,\left( -dt^2 + dq^2\right) \,+\, r_{\rm e}^2\, \left(d\theta^2 + \sin^2{\theta}d\phi^2 \right) ~.
\eeqa
Comparing to the previous metric in \eqref{eq:metric-1}, one has $dq = d\rho/(\rho^2/r_{\rm e}^2)$ and $e^{2 \sigma(q)} = \rho^2/r_{\rm e}^2 = r_{\rm e}^2/q^2$. The tetrads ${\rm e}^{A}_M$ are
\begin{align}
{\rm e}^{0}_{t}&=e^{\sigma}\,,\quad {\rm e}^{1}_{q}=e^{\sigma}\,,\quad {\rm e}^{2}_{\theta} =r_{\rm e}\,,\quad {\rm e}^{3}_{\phi}=r_{\rm e}\sin{\theta}~. 
\end{align}

The covariant derivative of the fermion is $D_{M}=\partial_{M} - \frac{i}{4}\,\omega_{M}^{AB}\,\Sigma_{AB}$ with the spin connection $\omega_{M}^{AB} = {\rm e}^A_N \nabla_M\,{\rm e}^{B N}$. The nonzero spin connections are $\omega^{01}_t = - \omega^{10}_t = d\sigma/dq$ and $\omega^{32}_\phi = - \omega^{23}_\phi = \cos{\theta}$. The Dirac operator is 
\beqa
\slashed{D} &\equiv& \gamma^{A}\,{\rm e}^{X}_{A}\,D_{X} \,\equiv\, \slashed{D}_{tq} \otimes \mathbb{I}_{2} + \frac{\sigma_{z}}{r_{\rm e}}\otimes \slashed{D}_{\theta\phi}  \nonumber \\
&=& e^{-\sigma}\Big{[}i \sigma_{x}\partial_{t} + \sigma_{y}\big{(}\partial_{q}+\frac{1}{2}\frac{d\sigma}{dq}\big{)}\Big{]}\otimes \mathbb{I}_{2} + \frac{\sigma_{z}}{r_{\rm e}}\otimes \Big{[}\sigma_{y}\frac{\partial_{\phi}}{\sin\theta} + \sigma_{x}\big{(}\partial_{\theta}+\frac{\cot{\theta}}{2}\big{)} \Big{]} 
~.
\eeqa
Using the notation of Ref.~\cite{spin-spherical-harmonics}, the Dirac operator in the $(\theta, \phi)$ can be written as
\beqa
\slashed{D}_{\theta\phi}  = 
\begin{pmatrix}
0 & - {_{\frac{1}{2}}}\eth^\prime\\
- {_{-\frac{1}{2}}}\eth & 0   
\end{pmatrix} ~,
\qquad \mbox{and} \qquad
\slashed{D}_{\theta\phi}\slashed{D}_{\theta\phi}  = 
\begin{pmatrix}
{_{\frac{1}{2}}}\eth^\prime \,{_{-\frac{1}{2}}}\eth & 0\\
0 & {_{-\frac{1}{2}}}\eth\, {_{\frac{1}{2}}}\eth^\prime
\end{pmatrix} ~,
\eeqa
where the operators ${_s}\eth\equiv - (\partial_\theta + \frac{i}{\sin{\theta}} \partial_\phi - s \,\cot{\theta})$ and ${_s}\eth^\prime \equiv - (\partial_\theta - \frac{i}{\sin{\theta}} \partial_\phi + s \,\cot{\theta})$. The spinor spherical harmonic functions ${_s} Y_{lm}(\theta, \phi)$ can be derived from the ordinary spherical functions by acting the operators ${_s}\eth$ and ${_s}\eth^\prime$ on them. They satisfy the conditions of $\eth^\prime \eth\,{_s} Y_{lm} = - (l - s)(l + s + 1)\,{_s} Y_{lm}$ and $\eth \eth^\prime\,{_s} Y_{lm} = - (l + s)(l - s + 1)\,{_s} Y_{lm}$~\cite{spin-spherical-harmonics}. Note that $l  \in \{\frac{1}{2},\frac{3}{2},\cdots\}$ and $m\in \{-l,l\}$~\cite{Abrikosov:2001nj}.

Defining two two-component vectors
\beqa
\mathcal{Y}_{lm}^{+}(\theta, \phi) \equiv
\begin{pmatrix}
{_{-\frac{1}{2}}} Y_{lm} \\
{_{\frac{1}{2}}} Y_{lm}
\end{pmatrix} \,,
\qquad 
\mathcal{Y}_{lm}^{-}(\theta, \phi) \equiv
\begin{pmatrix}
{_{-\frac{1}{2}}} Y_{lm} \\
-{_{\frac{1}{2}}} Y_{lm}
\end{pmatrix} ~,
\eeqa
the fermion wave functions can be expressed using separation of variables
\beqa
\Psi(t, q, \theta, \phi) = \psi_{lm}^+(t, q) \otimes \mathcal{Y}_{lm}^{+}(\theta, \phi) \, + \, \psi_{lm}^-(t, q) \otimes \mathcal{Y}_{lm}^{-}(\theta, \phi) ~. 
\eeqa
The AdS$_2$ two-dimensional equation of motion for the fermions are 
\beqa
&& \Big{(}\slashed{D}_{tq} \pm\frac{i \big{(}l+1/2\big{)}}{r_{\rm e}}\sigma_{z}\Big{)}\psi_{lm}^{\pm} (t, q) = 0 \,,\nonumber \\
&\mbox{or}& \slashed{D}_{tq} \slashed{D}_{tq} \psi^{\pm}_{lm}(t, q)  = \frac{1}{r^2_{\rm e}}\left( l + \frac{1}{2}\right)^2 \psi^{\pm}_{lm}(t, q) ~,
\eeqa
which means that the two-dimensional fermion masses are $|m| = (l  + 1/2)/r_{\rm e}$. Using the dictionary for fermions~\cite{Aharony:1999ti}, the corresponding operator scaling dimensions are $\Delta_{\mathcal{O}} = (d + 2\, |m|\,r_{\rm e})/2  = l + 1$ with $d=1$ and $l \ge 1/2$.

\subsection{Spin 1 (massless)}
\label{sec:spin-1-massless}
Our derivation in this section follows Ref.~\cite{Larsen:2014bqa}, where the 4-dimensional Lorentz gauge with $\xi = 1$ is used. We will use a general gauge parameter $\xi$, which makes the identification of physical states more transparent. 
We start with 4-dimensional Maxwell Lagrangian together with the gauge fixing term
\beq
\label{eq:lagrangian-massless}
\mathcal{L} = -\frac{1}{4} F_{MN}F^{MN} - \frac{1}{2 \xi} (\nabla_{\mu}a^{\mu} + \xi \nabla_{p}a^{p})^{2} ~,
\eeq
where $F_{MN}=\nabla_{M}a_{N}-\nabla_{N}a_{M}$ is the field tensor with $a_{M}$ as the gauge field, and $\nabla_{M}$ is the covariant derivative with $\nabla_{M}a_{N} = \partial_M a_N - \Gamma^{K}_{MN}a_K$. The gauge-fixing condition is 
\beq
\label{eq:gauge-fixing-massless}
\nabla_{\mu}a^{\mu}+\xi \nabla_{p}a^{p}=0 ~. 
\eeq

The gauge field is decomposed as 
\beqa
&a_{\mu}(x,y)& =  \sum_{l \geq 0, -l \le m \le l} \,b_{\mu}^{lm}(x)Y_{l}^{m}(y) \,, \nonumber \\
&a_{p}(x, y) & = \sum_{l \geq 1, -l \le m \le l} b^{lm}(x)\epsilon_{pq}\nabla^{q}Y_{l}^{m}(y) + \tilde{b}^{lm}(x) \nabla_{p}Y_{lm}(y) ~,
\eeqa
with $\epsilon_{\theta \phi}= \sin{\theta} \,\varepsilon_{\theta \phi}$ and $\varepsilon_{\theta \phi} = 1$ and  $\varepsilon_{\phi\theta } = -1$. Note that, $b_\mu^{lm}$ has $l$ start from 0, while $b^{lm}$ and $\tilde{b}^{lm}$ have $l$ start from 1.  Substituting the decomposed gauge field back into \eqref{eq:lagrangian-massless} and performing an integration in $(\theta, \phi)$, one has the 2-dimensional Lagrangian 
\begin{align}
\mathcal{L}_{2d} &= \frac{r_{\rm e}^2}{2}\,l(l+1)\, b^{lm}\left(\nabla_{A}^{2}-\frac{l(l+1)}{r_{\rm e}^{2}}\right) b^{lm} + \frac{r_{\rm e}^2}{2}\,l(l+1)\,\tilde{b}^{lm}\left(\nabla_{A}^{2}-\frac{\xi\,l(l+1)}{r_{\rm e}^{2}}\right)\tilde{b}^{lm} \nonumber\\
&\quad + \frac{r_{\rm e}^2}{2}\,b^{(lm)\mu}\Big{[}g_{2 \mu\nu}\nabla_{A}^{2}+\Big{(}\frac{1}{\xi} -1\Big{)}\nabla_{\mu}\nabla_{\nu} - g_{2 \mu\nu}\frac{l(l+1)}{r_{\rm e}^{2}} + g_{2 \mu\nu}\frac{1}{r_{\rm e}^{2}} \Big{]}b^{(lm)\nu} ~,
\end{align}
where $\nabla_{A}^{2}\equiv \nabla_{\mu}\nabla^{\mu}$. 

Next, we perform a Helmholtz-Hodge decomposition~\cite{Helmholtz1858,Hodge} of the 2-dimensional gauge field (see Ref.~\cite{hodge-decom} applications of this decomposition in many branches of science)
\begin{equation}
b^{lm}_{\mu} \,=\, b_{\mu \perp }^{lm} \,+\, b_{ \mu \parallel}^{lm} \,+ \,b_{\mu 0}^{lm} ~.
\end{equation}
Here, $b_{\mu \perp }^{lm}$ is the divergent-free (but not curl-free) component with $\nabla^{\mu}b_{\mu \perp }^{lm}=0$ ($\epsilon^{\mu\nu}\nabla_{\mu}b_{\nu \perp}^{lm}\neq 0$); $b_{\nu \parallel}^{lm}$ is the curl-free (but not divergent) component with $\epsilon^{\mu\nu}\nabla_{\mu}b_{\nu \parallel}^{lm}=0$ ($\nabla^{\mu}b_{\mu \parallel}^{lm}\neq0$); the last harmonic mode $b_{\mu 0}$ satisfies both divergent-free and curl-free conditions and also satisfies the equation of $(\nabla_{A}^{2}+1)b_{\mu 0}^{lm}=0$. 
To simplify the discussion, we further dualize the 2-dimensional vectors into scalars by $b_{\mu \perp}^{lm}=r_{\rm e}\,\epsilon_{\mu\nu}\nabla^{\nu}b_{\perp}^{lm}$, $b_{\mu \parallel}^{lm} = r_{\rm e}\,\nabla_{\mu}b_{\parallel}^{lm}$, and $b_{\mu 0}^{lm} =r_{\rm e}\, \nabla_{\mu}b_{0}^{lm}$. Because of the curl-free condition for the harmonic mode $b_{\mu 0}^{lm}$, its dual scalar field $b_{0}^{lm}$ satisfies the Laplacian equation $\nabla_{A}^{2}b_{0}^{lm} = 0$. In terms of those scalar fields, the initial 2-dimensional vector has the decomposition of 
\begin{equation}
r_{\rm e}^{-1}\,b^{lm}_{\mu} = \epsilon_{\mu\nu}\nabla^{\nu}b_{\perp}^{lm} + \nabla_{\mu}b_{\parallel}^{lm} + \nabla_{\mu}b_{0}^{lm} ~,
\end{equation}
with all fields having energy-dimension one. 

From $\mathcal{L}_{2d}$, the equations of motion for all fields are derived to be
\begin{align}
(\nabla_{A}^{2}-\frac{l(l+1)}{r_{\rm e}^{2}})\,b^{lm} &=0 \,, \, \qquad (\nabla_{A}^{2}-\frac{\xi\, l(l+1)}{r_{\rm e}^{2}})\,\tilde{b}^{lm} =0 ~, \\
(\nabla_{A}^{2}- \frac{l(l+1)}{r_{\rm e}^{2}})\,b^{lm}_{\perp} &=0 \,, \, \qquad
(\nabla_{A}^{2}- \frac{\xi\, l(l+1)}{r_{\rm e}^{2}})\,b^{lm}_{\parallel} =0 ~, \\
(\nabla_{A}^{2}- \frac{\xi\, l(l+1)}{r_{\rm e}^{2}})\,b^{lm}_{0} &=0  ~. \label{eq:harmonic}
\end{align}
Note that the on-shell equations of $b^{00}_{\perp}$ and $b^{00}_{\parallel}$ with $l = 0$  match to the condition for harmonic mode $\nabla_{A}^{2}b_{0}^{lm} = 0$. Therefore, only $l \ge 1$ modes exist for $b^{lm}_{\perp}$ and $b^{lm}_{\parallel}$, which is also true for $b^{lm}$ and $\tilde{b}^{lm}$. Also note that the equations of motion for $\tilde{b}^{lm}$, $b^{lm}_{\parallel}$ and $b^{lm}_{0}$ with $l \ge 1$ are all gauge-parameter dependent, so the remaining physical degrees of freedom are $b^{lm}$ and $b^{lm}_\perp$ with $l \ge 1$ and $b^{00}_0$. The harmonic mode $b^{00}_0$ is known as a non-renormalizable boundary mode~\cite{Larsen:2014bqa}. Therefore, the propagating 2-dimensional modes $b^{lm}$ and $b^{lm}_\perp$ have masses of $m^2 = l (l + 1)/r_{\rm e}^2$ with $l \ge 1$. The corresponding operators have the scaling dimension of $\Delta_{\mathcal{O}} = l + 1$ with $l \ge 1$. 

One could also use the gauge condition plus gauge transformation to identify the on-shell physical degrees of freedom. The gauge condition has $\nabla_{\mu}a^{\mu}+\xi \nabla_{p}a^{p}=0$. Using the field decompositions and the equation of motion for $b^{lm}_{\parallel}$, one has $b^{lm}_{\parallel} = \tilde{b}^{lm}$. For the gauge transformation $a_M \rightarrow a_M + \nabla_M \Lambda$ with $\Lambda(x, y) = \lambda^{lm}(x) Y^m_l(y)$, one has $\tilde{b}^{lm}\rightarrow \tilde{b}^{lm} +\lambda^{lm}$ and $b_{\mu \parallel}^{lm}\rightarrow b_{\mu \parallel}^{lm} + \nabla_{\mu}\lambda^{lm}$. Because $b_{\mu \parallel}^{lm} = \nabla_{\mu}b_{\parallel}^{lm}$, one can identify $\tilde{b}^{lm} = b_{\parallel}^{lm} = \lambda^{lm}$ and have the configuration gauge equivalent to vacuum. For the harmonic modes $b^{lm}_0$, the harmonic condition $\nabla_{A}^{2}b_{0}^{lm} = 0$ and the equation of motion in \eqref{eq:harmonic} simply implies that only $l = 0$ is physical.

\subsection{Spin 1 (massive)}
\label{sec:spin-1-massive}
Since the SM has spontaneous breaking of the electroweak gauge symmetry, one also needs to consider additional degrees of freedom including the Higgs boson and longitudinal and/or Goldstone Boson (GB) degrees of the massive gauge bosons. Here, we use a simpler $U(1)$ gauge model to identify the physical propagating modes. A short summary result is that the Higgs boson has one tower of KK modes with $m^2 = l(l + 1)/r_{\rm e}^2 + m_h^2$ with $l \geq 0$ and the gauge boson plus the GB have three towers of KK modes with $m^2 = l(l + 1)/r_{\rm e}^2 + g^2\,v^2$ with $l \geq 1$ plus one $l=0$ state with $m^2 = g^2\,v^2$. In the limit of $1/r_{\rm e} \gg m_h, g v$, the corresponding operators have $\Delta_{\mathcal{O}} = l + 1$ with $l \ge 0$ related to the Higgs boson and $\Delta_{\mathcal{O}} \approx l + 1$ with $l \ge 1$ and a multiplicity of 3 and $\Delta_{\mathcal{O}} \approx 1$ with a single degree of freedom related to the gauge boson plus GB.  Applying this counting to the SM electroweak sector and concentrating on the operators with the lowest scaling dimension, we have four operators with a scaling dimension $\Delta_{\mathcal{O}} \approx 1$ with the corresponding field masses: $m_h =125$~GeV, $M_{W^\pm} = 80.4$~GeV  and $M_Z = 91.2$~GeV.

The 4-dimensional Lagrangian including the Lorentz-invariant gauge-fixing term is
\beqa
\mathcal{L}  = -\frac{1}{4} F_{MN}F^{MN} + D^M \Phi^\dagger D_M \Phi + V(\Phi^\dagger \Phi) - \frac{1}{2} ( \partial_M A^M  + g\,v\,\varphi)^{2} ~,
\eeqa
with $D_M \Phi = \partial_M \Phi + i \,g \, A_M \, \Phi$ and $\Phi =(v + h + i\,\varphi)/\sqrt{2}$. Here, $h$ is the Higgs boson; $\varphi$ is the Goldstone boson; $v$ is the vacuum expectation value of $\Phi$.  The scalar potential $V(\Phi^\dagger \Phi)$ contains the positive Higgs boson mass term $\frac{1}{2}\, m_h^2\, h^2$ and no mass term for $\varphi$. Decomposing 4-dimensional fields in the similar way as in Sections~\ref{sec:spin-0} and \ref{sec:spin-1-massless}, we have the equations of motion for the 2-dimensional fields as
\begin{align}
& (\nabla_{A}^{2}-\frac{l(l+1)}{r_{\rm e}^{2}} - m_h^2 )\,h^{lm}  = 0  \,,  &  (\nabla_{A}^{2}-\frac{l(l+1)}{r_{\rm e}^{2}} - g^2\,v^2 )\,\varphi^{lm}  = 0 ~,  \\
&(\nabla_{A}^{2}-\frac{l(l+1)}{r_{\rm e}^{2}} - g^2\,v^2)\,b^{lm} =0 \,, \, & (\nabla_{A}^{2}-\frac{ l(l+1)}{r_{\rm e}^{2}} - g^2\,v^2)\,\tilde{b}^{lm} =0 ~, \\
& (\nabla_{A}^{2}- \frac{l(l+1)}{r_{\rm e}^{2}} - g^2\,v^2)\,b^{lm}_{\perp} =0 \,, \,
&(\nabla_{A}^{2}- \frac{l(l+1)}{r_{\rm e}^{2}} - g^2\,v^2)\,b^{lm}_{\parallel} =0 ~, \\
\label{eq:b0-eom}
&(\nabla_{A}^{2}- \frac{l(l+1)}{r_{\rm e}^{2}} - g^2\,v^2)\,b^{lm}_{0} =0 ~.
\end{align}
Here, $h^{lm}$, $\varphi^{lm}$, $b^{lm}_{\perp}$, $b^{lm}_{\parallel}$ and $b^{lm}_{0}$ have $l \ge 0$, while $b^{lm}$ and $\tilde{b}^{lm}$ have $l \ge 1$. 

For $l \ge 1$ modes, we can use the gauge-fixing condition and the gauge transformation to identify one physical KK tower for three fields $\varphi^{lm}$, $\tilde{b}^{lm}$ and $b^{lm}_{\parallel}$. Starting from the gauge fixing condition $\partial_M A^M + g \, v\, \varphi = 0$ and using the equation of motion for $b^{lm}_{\parallel}$, one has 
\beqa
\label{eq:gauge-fixing}
\left(\frac{l(l+1)}{r_{\rm e}^{2}} + g^2\,v^2 \right) \,b^{lm}_{\parallel}  - \frac{l(l+1)}{r_{\rm e}^{2}}\,\tilde{b}^{lm} + \frac{g\,v}{r_{\rm e}}\, \varphi^{lm}  = 0 ~.
\eeqa
The gauge transformation has $a_M \rightarrow a_M + \partial_M \Lambda$ and $\varphi \rightarrow \varphi - g\,v\,\Lambda$. In terms of the 2-dimensional fields, one has 
\begin{align}
\label{eq:gauge-transformation}
& b^{lm}_{\parallel} \rightarrow b^{lm}_{\parallel} + r_{\rm e}^{-1}\,  \lambda^{lm} ~,  
\qquad \tilde{b}^{lm} \rightarrow \tilde{b}^{lm} + r_{\rm e}^{-1}\,\lambda^{lm} ~, 
\qquad \varphi^{lm} \rightarrow  \varphi^{lm} - g\,v\,\lambda^{lm} ~. 
\end{align}
This means that the equally linear combination of $[r_{\rm e}\,b^{lm}_{\parallel}, r_{\rm e}\,\tilde{b}^{lm}, -\varphi^{lm}/(g\,v)]$ is equivalent to vacuum up to gauge transformation. After subtracting this linear combination and the one constrained by the gauge fixing condition in \eqref{eq:gauge-fixing}, there is one physical propagating mode with mass $m^2 = l(l+1)/r_{\rm e}^2 + g^2 v^2$. There are two additional modes, $b^{lm}$ and $b^{lm}_\perp$, with $l \ge 1$ and the same mass. Note that $b^{lm}_0$ vanishes after using the equation of motion and the harmonic mode constraint. So, for $l\ge 1$, we have three towers of KK modes with the same mass. 

For $l = 0$, we first note that $b^{00}_\parallel$ and $\varphi^{00}$ are gauge equivalent to vacuum after using \eqref{eq:gauge-fixing} and \eqref{eq:gauge-transformation}, so they are not physical. The mode $b^{00}_0$ vanishes from its equation of motion in \eqref{eq:b0-eom} and the harmonic condition $\nabla_{A}^{2}b_{0}^{00} = 0$. We are left with $b^{00}_\perp$ with an equation of motion $(\nabla_{A}^{2} - g^2 v^2)b^{00}_\perp = 0$ and not in conflict with the condition of independence of the harmonic mode: $\nabla_{A}^{2} b^{00}_\perp \neq 0$. 

Altogether we have the physical propagating modes: $h^{lm}$, $b^{lm}_\perp$ with $l \ge 0$ and $b^{lm}$ plus one linear combination of $(\varphi^{lm}, \tilde{b}^{lm}, b^{lm}_\parallel)$ with $l \ge 1$. The two $l=0$ modes, $h^{00}$ and $b^{00}_\perp$, have their mass-squared as $m_h^2$ and $g^2 v^2$, respectively. Their corresponding CFT$_1$ operator scaling dimensions are $\Delta_{\mathcal{O}} \approx 1 + m_h^2\, r_{\rm e}^2$ and $1 + g^2\,v^2\, r_{\rm e}^2$ in the limit of $m_h, g\,v \ll r_{\rm e}^{-1}$. Note that in the small gauge coupling limit $g\rightarrow 0$, the complex scalar and the massless gauge boson decouple from each other. The propagating physical modes should just contain the massive scalar Higgs boson and the massless scalar Goldstone boson together with massless spin-1 gauge boson.

\subsection{Spin 2}
For the massless spin-2 case, a similar analysis as the massless spin-1 case has been performed in Ref.~\cite{Larsen:2014bqa}. We do not repeat it here. The physical propagating degrees of freedom are two KK towers of scalar fields with $m^{2}=l(l+1)/r_{\rm e}^{2}$ with $l \geq 2$, which implies the corresponding towers of operators with a scaling dimension of $\Delta_{\mathcal{O}}= l + 1$ with $l\geq 2$.


\providecommand{\href}[2]{#2}\begingroup\raggedright\endgroup

\end{document}